\documentclass[preprint,12pt]{elsarticle}

\usepackage{amssymb}

\begin{document}

\begin{frontmatter}



\title{Photoalignment at the nematic liquid crystal - polymer interface: experimental evidence of three-dimensional reorientation}


\author{Tibor T\'oth-Katona\corref{1}}
\cortext[1]{Corresponding author.}
\ead{tothkatona.tibor@wigner.mta.hu}

\author{Istv\'an J\'anossy}

\address{Institute for Solid State Physics and Optics, Wigner Research Centre for Physics,
Hungarian Academy of Sciences, P.O. Box 49, H-1525 Budapest, Hungary.}

\begin{abstract}
We provide experimental evidence that photoalignment at the nematic liquid crystal (NLC) - polymer interface can not be simply considered as a two-dimensional process. Moreover, our experiments clearly indicate that the photoaligning process does not depend on the individual properties of the NLC material and those of the interfacing polymer exclusively. According to our measurements, the polymer and the NLC layer "sense-each-other", i.e., the polymer-liquid crystal interface should be regarded as a coupled system, where the two components mutually influence each other. Furthermore, we show that the temperature induced anchoring transition also has to be taken into account for the complete description of the photoalignment mechanism.

\end{abstract}


\end{frontmatter}


\section{Introduction}
\label{Intro}

The interaction of liquid crystals with solid substrates is a particularly significant research area. The problem is important for both the basic research and for various applications, because the proper alignment of the molecules at the boundary of a liquid crystal cell is the key factor for correct operation of devices based on LCs. Although standard methods have been developed in the last decades to ensure the required orientation of liquid crystals (like mechanical rubbing of polyimide layers spin-coated on the solid substrate), there is a continuous search for new methods of alignment. One of the most stimulating alternative method is the so-called photoalignment of nematic liquid crystals, discovered in the early 1990-s \cite{Gibbons1991,Dyadyusha1992,Gibbons1995}. Moreover, photoalignment can be used not only to ensure the desired orientation in LC devices, but opens up the possibility to reorient the liquid crystal director through light irradiation. The exceptional advantage of the orientation control by light lies in its contactless nature \cite{Ube2017}.

Photoalignment is most often realized exploiting trans-cis (E/Z) isomerization of azo dyes \cite{Rau1990}. The azo dyes are either coated on the substrate as a molecular monolayer \cite{Aoki1992,Yi2009,Janossy2011,Janossy2014,Janossy2014a}, or embedded in a polymer film
\cite{Gibbons1991,Gibbons1995,Ichimura1989,Ichimura1993,Akiyama1995,Janossy1999,Janossy2001,Palffy2002}. In a typical experiment, a liquid-crystal (LC) sandwich cell is constructed with one photosensitive substrate and one traditionally prepared reference plate. In one scheme, the trans (E) configuration of the given azo substance induces homeotropic orientation on the photosensitive plate (the liquid crystal molecules are aligned perpendicularly to the substrate), while the cis (Z) isomer induces planar one (the liquid crystal is oriented in the plane of the substrate). In this case, the light irradiation creates cis isomers, triggering a homeotropic-to-planar transition ‒ often called 'out-of-plane alignment' photocontrol \cite{Aoki1992,Ichimura1993,Kawanishi1991}. More often, with other azo substances, the liquid crystal molecules remain in the plane of the substrate and the azimuthal angle of the director can be controlled with polarized light, therefore creating a twisted nematic cell. As a rule, after irradiation the director becomes perpendicular to the light polarization direction, and is referred to as the 'in-plane alignment' photocontrol \cite{Gibbons1991,Gibbons1995,Ichimura1993,Akiyama1995,Palffy2002}. The mechanisms of the photocontrol, as well as the most commonly used photosensitive materials are well summarized in a review on photoalignment of liquid crystal systems \cite{Ichimura2000}.

Hereby we present observations that have far-reaching consequences regarding the photoalignment process, and can change the concept of its mechanism profoundly. First, we will show that the standard two-dimensional description of photoalignment is unsuitable in a number of cases. Second, we will provide evidence that the nematic liquid crystal does not play a
purely passive part in the photoalignment process, in contrast to what is assumed in most papers. Instead, the polymer-liquid crystal
interface should be regarded as a coupled system, where the two components mutually influence each other \cite{Janossy2001}.

\section{Experimental details}

\label{Exp}

\subsection{Sample preparation}

To obtain the photosensitive substrate, the following procedure has been employed.
The polymer polymethyl-methacrilate functionalized with the azo-dye Disperse Red 1 (pDR1, for the chemical structure see Fig. 1b of \cite{Janossy2018}) has been dissolved in toluene (concentration of pDR1: $2 wt\%$) and spin-coated on glass plates (some of them having $\mathrm {SnO_2}$ conducting layer serving as an electrode). Spin-coating has been performed with Polos SPIN150i at 800rpm for 5sec, and then at 3000rpm for 30 seconds (all with spin acceleration of $\pm$1000 rpm/sec). The spin-coated substrates has been baked in an oven for about 2 hours at $140^{\circ}$C.

Rubbed polyimide coated slides from E.H.C. Co. (Japan) have been used as reference plates. The thickness of the assembled LC cells has been measured by interferometric method, and were found in the range of $7-50\mu$m. The LC cells have been filled either with 4-cyano-4'-penthylbiphenyl (5CB), with LC mixture E7, or with mixture E63 (both from Merck). Before filling in the material, the cell was illuminated with white light, polarized perpendicularly to the rubbing direction on the reference plate. This procedure ensured a good quality planar initial alignment of the
nematic liquid crystal. The nematic-to-isotropic phase transition temperature $T_{NI}$ has been determined for all LC cells, and typical values of
$34^{\circ}$C, $60^{\circ}$C and $83^{\circ}$C have been found for 5CB, E7 and E63, respectively.

\subsection{Experimental setup}

Measurements on the photoalignment and photo-reorientation have been performed on an improved version of the pump-probe optical setup developed in our laboratory recently, and described in details in Ref. \cite{Janossy2018}.
The polarized pump beam from a DPSS laser ($25$mW, $\lambda=457$nm) entered the cell from the photosensitive side, defocussed to a spot size of few mm (much larger than the diameter of the probe beam). The polarized probe beam from a He-Ne laser ($5$mW, $\lambda=633$nm) was sent through the cell, entering it at the reference plate. Behind the sample the probe beam was sent through a rotating polarizer and its intensity was detected by a photodiode; the signal was connected to a lock-in amplifier. The setup provides the phase and the amplitude of the probe beam transmitted through the sample. In order to determine the induced twist angle, the probe beam was polarized parallel to the rubbing direction; the twist angle is given by the phase of the signal. To detect the zenithal reorientation, the probe beam was polarized at $45^{\circ}$ from the rubbing direction and the amplitude of the signal was measured. For the determination of the initial pretilt angle at the photosensitive substrate an additional electric-field has been subjected to the sample in a standard measurement setup for the electric-field-induced Fr\'eedericksz transition, in which the sample is placed between crossed polarizers (polarization direction of both enclosing $45^{\circ}$ with the initial director {\bf n}), and the transmitted light intensity from a He-Ne laser is measured during the transition.

\section{Results}
\label{Results}

\subsection{Azimuthal (in-plane) photoalignment}

We start with the results on the in-plane (azimuthal) photoalignment. In these measurements the polarization of both the pump and the probe beam has been set parallel with the initial {\bf n}, and the phase of the probe beam has been measured. The temperature of the samples has been varied from room temperature up to $T_{NI}$. Under these conditions the so called 'in-plane alignment' photocontrol is expected, so that the pump beam induces a twist deformation at the photosensitive substrate resulting in a twisted LC cell from the initially planar one. In the case of a perfect azimuthal reorientation the twist angle should be $90^{\circ}$.

\begin{figure}
\begin{center}
{\bf (a)}\includegraphics[width=23pc]{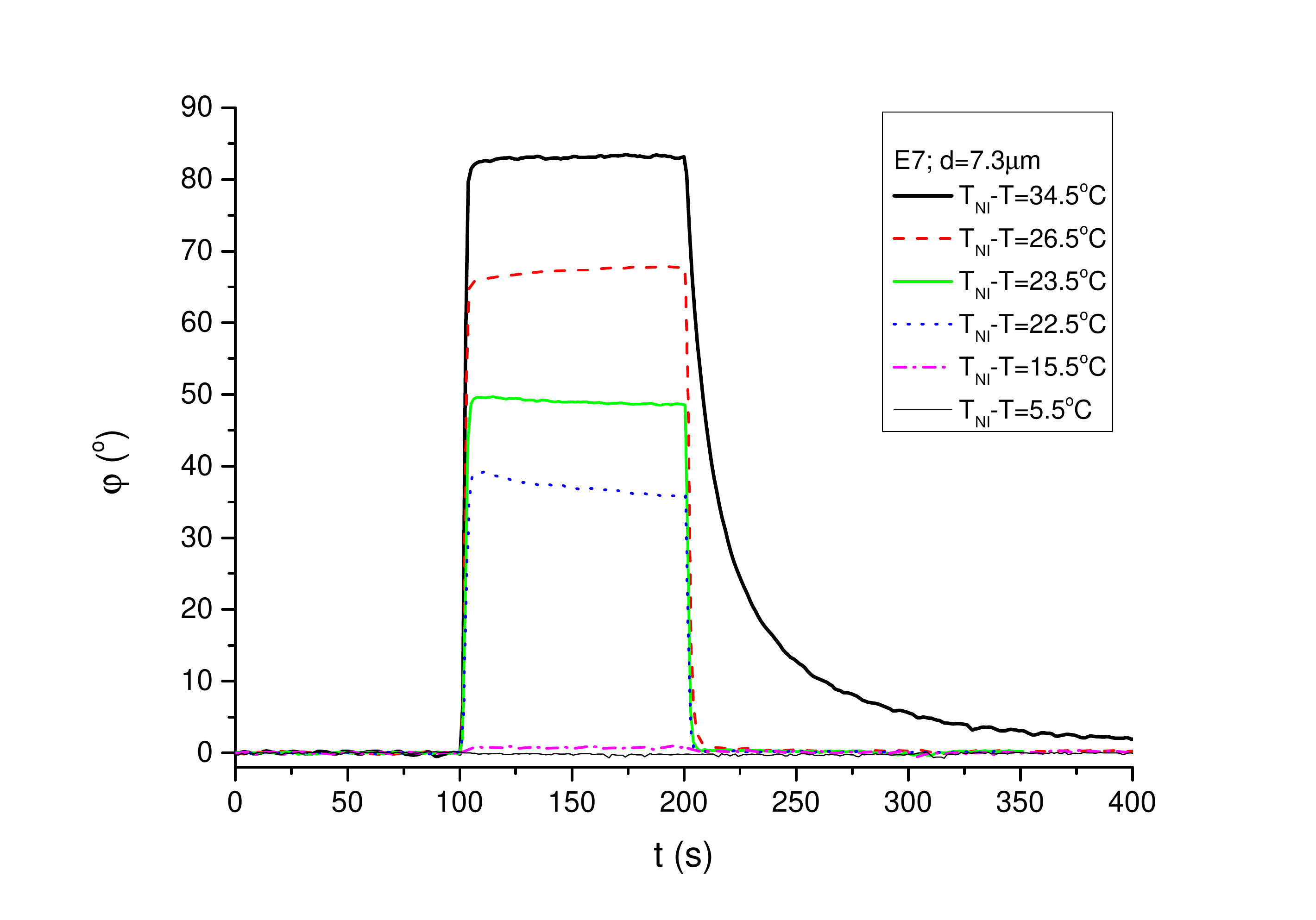} {\bf (b)}\includegraphics[width=23pc]{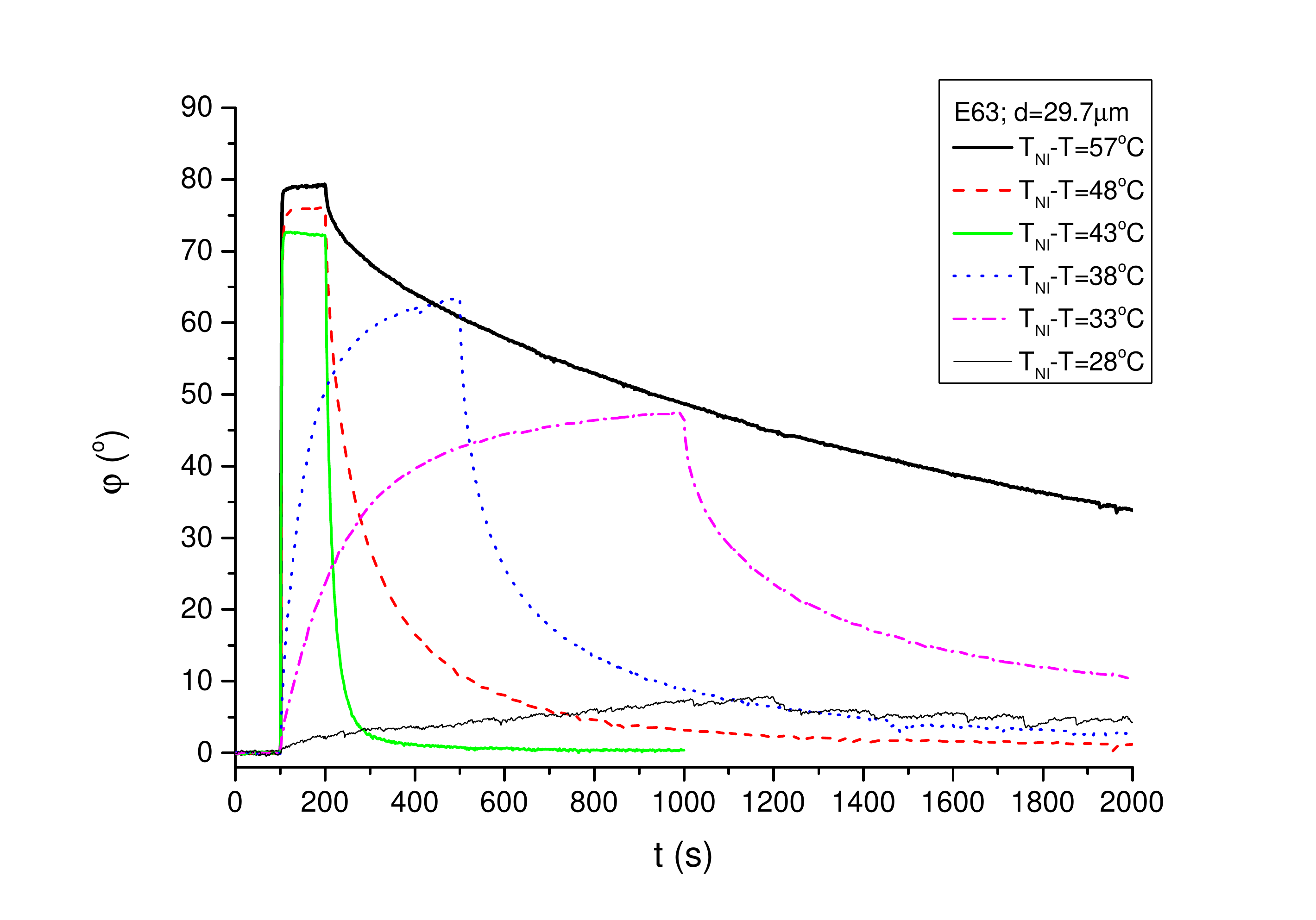} {\bf (c)}\includegraphics[width=23pc]{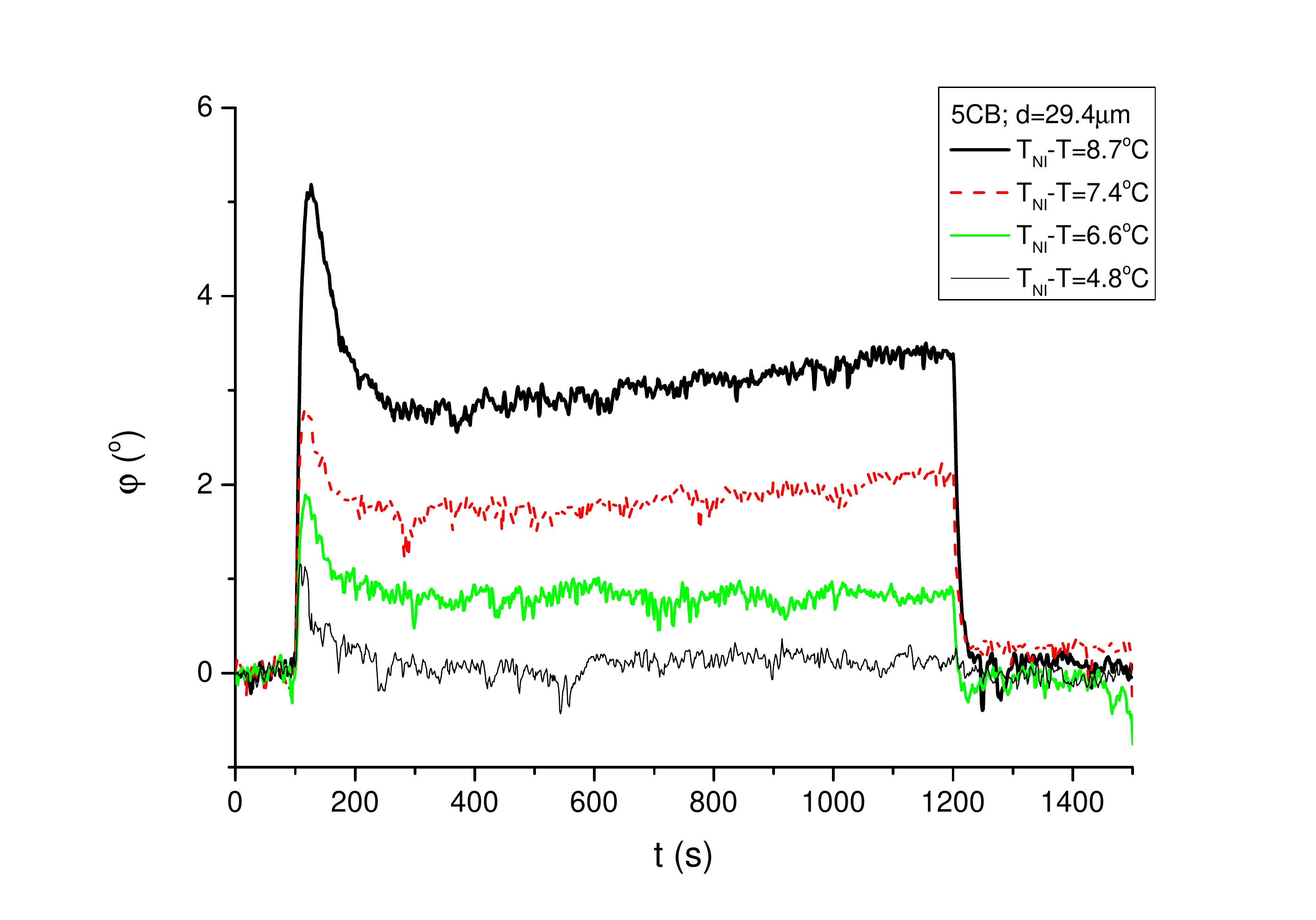}
\caption{Azimuthal photo-reorientation angle $\varphi$ in time, measured at different temperatures in cells filled with NLCs: (a) E7, (b) E63, and (c) 5CB. The pump-beam was switched on at $t=100$s, and switched off at a time ranging from 200s to 1200s, depending on the photo-reorientation dynamics.} \label{Toth_Fig1}
\end{center}
\end{figure}

Figure \ref{Toth_Fig1} shows the temporal evolution of the photoinduced twist angle $\varphi$ for LC cells filled with E7 (a), E63 (b), and 5CB (c), measured at different temperatures
$\Delta T = T_{NI}-T$.
The pump-beam has been switched on at $t=100$s in all measurements, and switched off somewhere in the range $t=200 - 1200$s, depending on the dynamics of the photoalignment. As one sees, for E7 and E63 at low (room) temperature the azimuthal twist deformation saturates at $\varphi \geq 80^\circ$ [see Fig.~\ref{Toth_Fig1}(a) and (b)], which is close to the complete reorientation $\varphi = 90^\circ$. With the increase of the temperature, however, the twist angle gradually decreases, and vanishes far below $T_{NI}$. Moreover, for 5CB no significant twist deformation has been detected -- $\varphi <4^{\circ}$ has been measured even at the room temperature (Fig.~\ref{Toth_Fig1}(c)). Besides the decrease of the twist angle with the increase of the temperature, from Fig.~\ref{Toth_Fig1} -- especially from subfigure (b) -- one can also see that the photoalignment process slows down, while (after switching off the pump beam) the back relaxation speeds up with the increase of the temperature.

\begin{figure}
\begin{center}
\includegraphics[width=23pc]{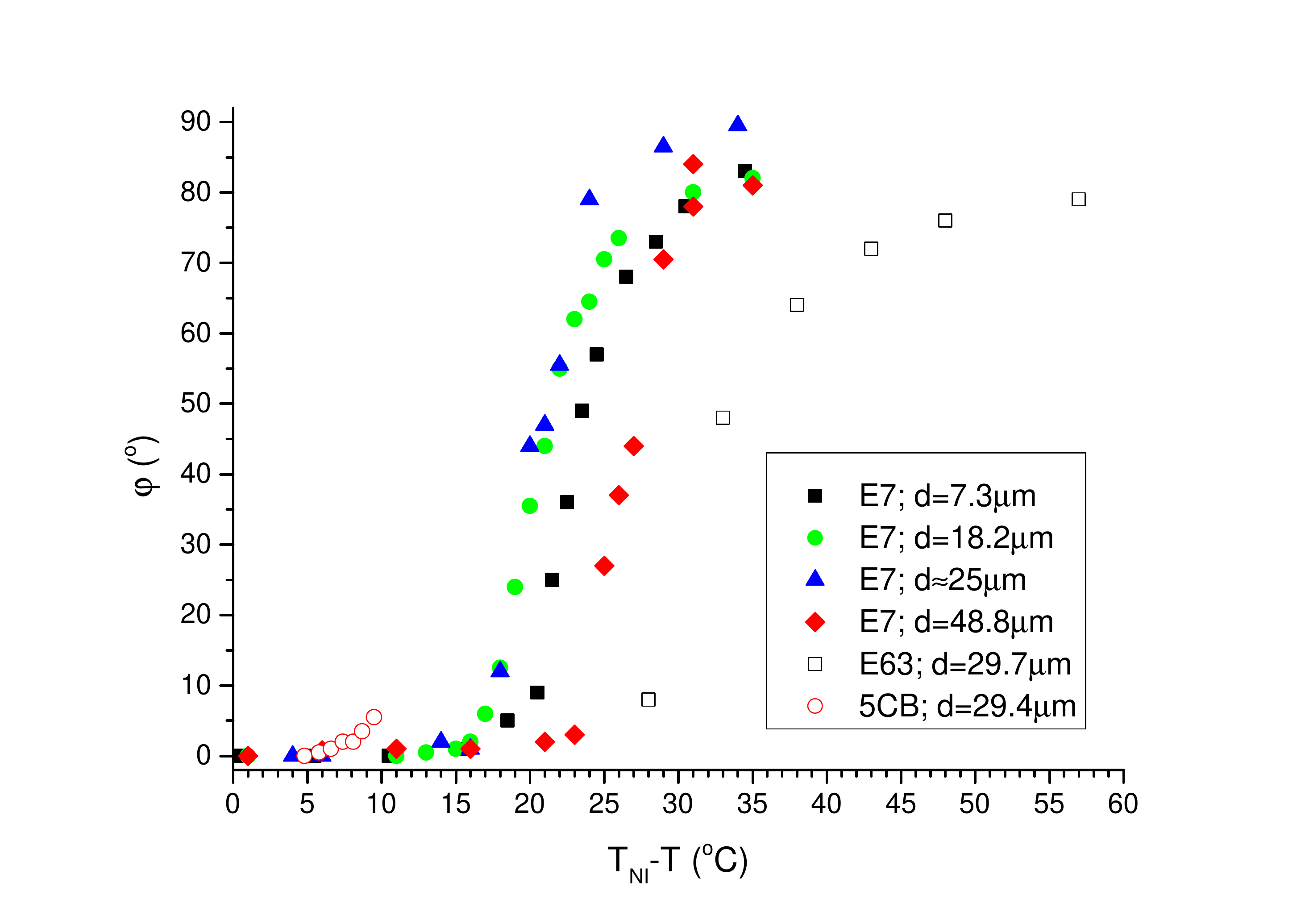}
\caption{Temperature dependence of the azimuthal photo-reorientation angle $\varphi$ measured in cells of thickness $d$ as indicated in the legend, and filled with various NLCs (E7, E63 and 5CB).} \label{Toth_Fig2}
\end{center}
\end{figure}

The temperature dependence of the saturated twist angle for 5CB, E7 and E63 is plotted in Fig.~\ref{Toth_Fig2}. For all NLCs a sudden decrease of $\varphi$ has been observed far below $T_{NI}$: for E63 in the temperature range $T_{NI}-T = 30-35^{\circ}$C, for E7 in the range of $T_{NI}-T = 20-25^{\circ}$C, and for 5CB the range of $T_{NI}-T > 10^{\circ}$C can be deduced from the measurements (the present experimental setup does not allow for measurements below the ambient temperature). In Fig.~\ref{Toth_Fig2} results for E7 measured in LC cells having different thickness $d$ are also shown. No systematic $d$ dependence of $\varphi$ is observed. It is worth to note, however, that according to measurements the higher is the $T_{NI}$ of the NLC, the sudden decrease of $\varphi$ occurs at larger values of the relative temperature $\Delta T=T_{NI}-T$. In summary, in all samples a broad temperature range of the nematic phase is detected in which practically no azimuthal photo-reorientation is observed, i.e., $\varphi \approx 0$.

\subsection{Temperature-induced zenithal reorientation}

In this section we investigate the temperature dependence of the pretilt angle $\theta$ at the photosensitive substrate. We note that the pretilt angle at the reference plate (rubbed polyimide) has been found temperature independent over a wide temperature range for various nematic liquid crystals \cite{Myrvold1995}. In particular, for the type of the polyimide substrate used in our experiments (manufactured by E.H.C. Co., Japan) the pretilt is estimated in the range $0<\theta <1^{\circ}$ \cite{Urbanski2014}. We have used the electric field induced Fr\'eedericksz transition \cite{Freed1927} in order to estimate the pretilt angle $\theta$ at the photosensitive substrate. Namely, NLCs with positive dielectic anisotropy $\Delta \varepsilon = \varepsilon_{\parallel}- \varepsilon_{\perp}$ ($\varepsilon_{\parallel}$ and $\varepsilon_{\perp}$
are the dielectic permittivities parallel with-, and perpendicular to {\bf n}, respectively) and with planar initial alignment, undergo an orientational (Fr\'eedericksz) transition upon application of an electric field at a threshold voltage $U_F = \pi \sqrt{K_{11}/(\varepsilon_0 \Delta \varepsilon)}$, where $K_{11}$ is the splay elastic constant. We note that $U_F$ is a material parameter, which does not depend on geometrical factors. However, the threshold character of the electric-field induced reorientation is connected with the strictly planar alignment. If the director configuration deviates from the perfect planar alignment, i.e., there is a pretilt, the threshold-like behaviour of the reorientation is smoothed out.  The birefringence vs. applied voltage curves around the Fr\'eedericksz threshold voltage exhibit an inflexion point. The curves are extremly sensitive to the pretilt angle, especially below and around  $U_F$ \cite{Urbanski2014}. Taking this fact into account, we have measured the birefringence change as a function of the applied voltage, and compared the results with numerical calculations in which only the pretilt $\theta$ at the photosensitive substrate entered as a free parameter. The measurements were performed from the room temperature to $T_{NI}$.

As already mentioned, measurements on the Fr\'eedericksz transition have been performed in a standard setup with crossed polarizers for 5CB and E7, for which the temperature dependence of all relevant material parameters is available in the literature. At each temperature the transmitted light intensity is measured with voltage steps of 0.02V in the range from 0.04V to 7V, in an increasing and decreasing voltage cycle, with waiting time of 1s between each step.

Numerical calculations are based on equations given in \cite{Deuling1972}, to which we have included the influence of the pretilt angle at both bounding substrates of the LC cell.
The sample thickness $d$ and the phase transition temperature $T_{NI}$ have been measured for each LC cell ($d=17.6\mu$m and $20.2\mu$m, $T_{NI}=34^{\circ}$C and $59^{\circ}$C, for 5CB and E7, respectively), while the wavelength of the probe beam ($\lambda = 633$nm) is known. The temperature dependence of the relevant material parameters has been taken from the literature as follows: splay and bend elastic constants [$K_{11}(T)$ and $K_{33}(T)$, respectively] for E7 from \cite{Raynes1979}, while for 5CB from \cite{Bogi2001} which is in excellent agreement with another independent measurement \cite{Bradshaw1985}; the ordinary and extraordinary refractive indices [$n_o(T)$ and $n_e(T)$, respectively] from \cite{Li2005} for E7, and from \cite{Karat1976} for 5CB; the dielectric permittivities parallel with-, and perpendicular to {\bf n} [$\varepsilon_{\parallel}$ and $\varepsilon_{\perp}$, respectively] from \cite{Raynes1979} for E7 and from \cite{Bogi2001} for 5CB. In the calculations, in agreement with \cite{Urbanski2014}, a temperature independent pretilt angle of $0.3^{\circ}$ and $1^{\circ}$ has been chosen at the reference plate for E7 and 5CB, respectively. We will see in the followings that such a somewhat arbitrary choice of the pretilt at the interface with the rubbed polyimide layer (in the range of $0 < \theta_0 \leq 1^{\circ}$) is not influencing the main  results on the pretilt angle $\theta$ at the photosensitive substrate.

After the choice of parameters as described above, only the pretilt angle $\theta$ at the interface with pDR1 remains the free fit parameter in numerical calculations to compare with the  experimental results. In the calculations, the measured sample thickness has also been slightly adjusted (within about $\pm 5\%$ compared to the measured one), in order to bring together the measured and calculated normalized light intensities together at low voltages, far below $U_{F}$. In this sense $d$ can be regarded as a quasi-fit parameter, however, such a small variation of $d$ may also originate from the error of the interferometric method, or from spatial thickness variations within the LC cell.

\begin{figure}
\begin{center}
\includegraphics[width=23pc]{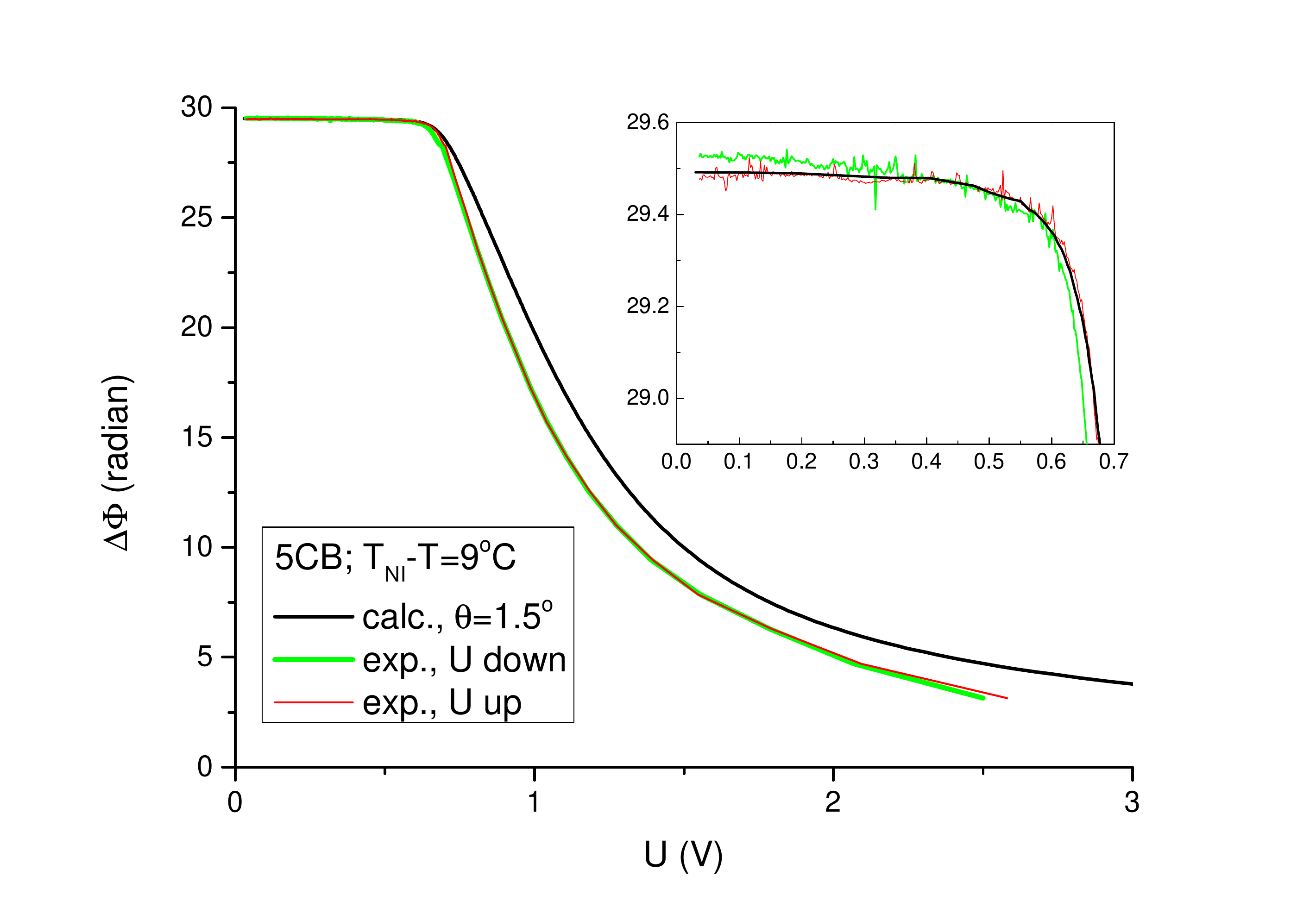}
\caption{Comparison of the calculated total phase shift with the measured one in an increasing/decreasing voltage cycle ($U$ up/$U$ down cycle) for a 5CB sample at $\Delta T= T_{NI}-T = 9^{\circ}$C. Inset: the blowup of the voltage range below $U_{F}=0.72$V.} \label{Toth_Fig3}
\end{center}
\end{figure}

The total phase shift $\Delta \Phi$ can be calculated from the light intensity variations observed while the sample undergoes the Fr\'eedericksz transition \cite{Bogi2001}. In Fig. \ref{Toth_Fig3} we give a representative example of the voltage dependence of $\Delta \Phi$ measured in a 5CB sample in increasing ($U$ up) and decreasing ($U$ down) voltage steps, together with the calculated $\Delta \Phi (U)$ curve for a pretilt of $\theta = 1.5^{\circ}$. Fig. \ref{Toth_Fig3} is representative in the sense that reflects in general the characteristics observed both for 5CB and E7 LCs in the whole temperature range of the nematic phase. The measured $\Delta \Phi (U)$ deviates from the calculated one at higher voltages: in the experiments the Fr\'eedericksz transition ends at a somewhat lower voltage than it is predicted by the calculations. This systematic deviation (presumably originating from the values of material parameter(s) taken from the literature) occurs at all temperatures, both for 5CB and for E7, and could not be eliminated by adjusting the fit parameter $\theta$. Below $U_{F}$ (where the influence of $\theta$ is the most significant), however, an excellent agreement between the experiments and the calculations could be reached just by fitting the value of $\theta$ -- see the inset in Fig. \ref{Toth_Fig3}.

\begin{figure}
\begin{center}
\includegraphics[width=23pc]{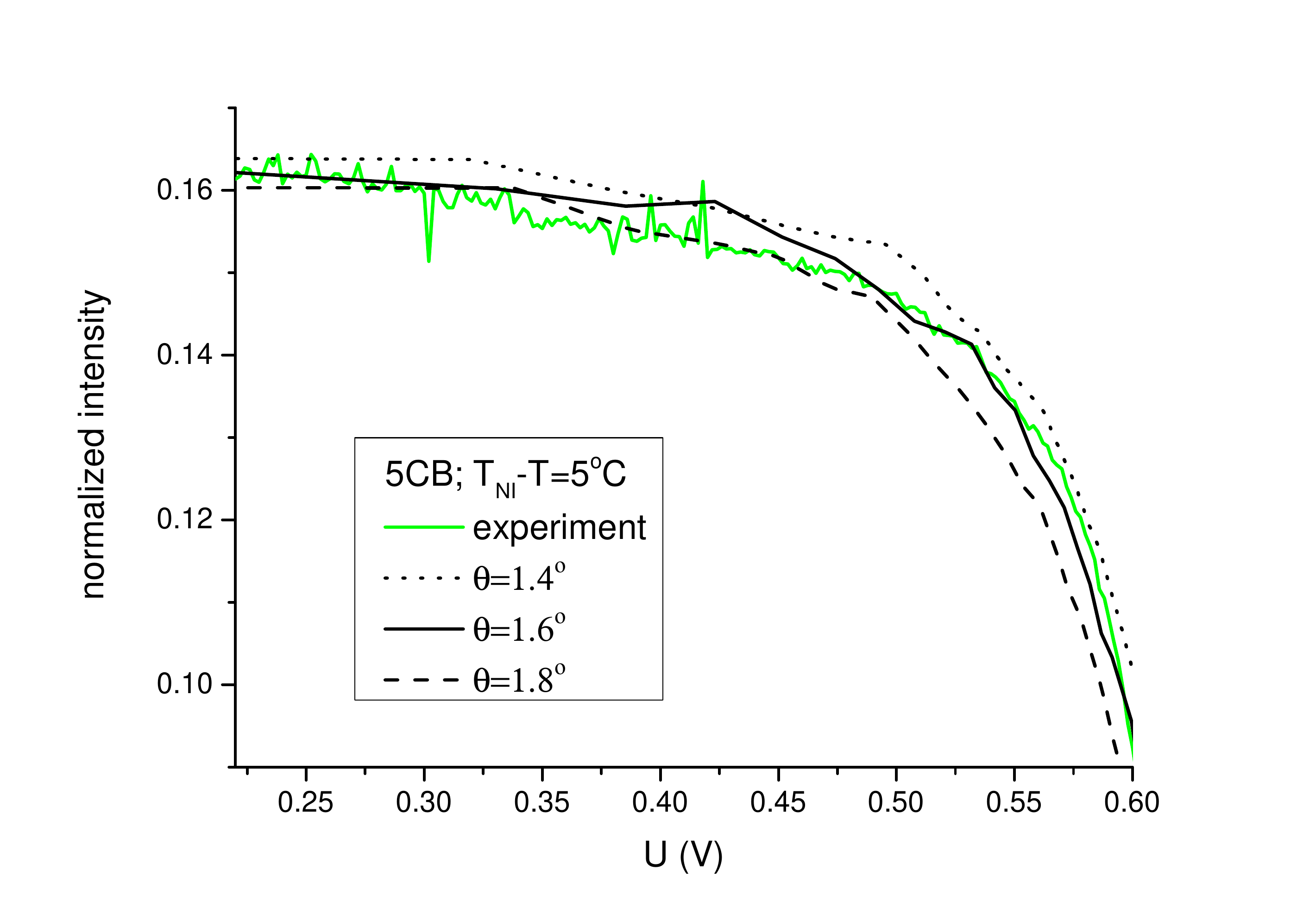}
\caption{Voltage dependence of the normalized light intensity $I$ below $U_{F}=0.68$V, measured in 5CB at $\Delta T = 5^{\circ}$C and calculated for three slightly different pretilt angles $\theta$.} \label{Toth_Fig4}
\end{center}
\end{figure}

 The sensitivity of the onset of the Fr\'eedericksz transition to the variation of the fit parameter $\theta$ is illustrated in Fig. \ref{Toth_Fig4}, where the voltage dependence of the normalized light intensity $I$ is shown below $U_{F}$ for the measurement with increasing voltage steps and for calculations with slightly different pretilt angles ($1.4^{\circ}$, $1.6^{\circ}$ and $1.8^{\circ}$) in 5CB at $\Delta T = 5^{\circ}$C. As one sees, even such a small variation of $\theta$ causes a considerable change: $\theta=1.4^{\circ}$ seems to give give an overestimate of the experimental curve, while $\theta=1.8^{\circ}$ underestimates it. The best fit to experimental data is obtained for $\theta=1.6^{\circ}$. In general, for both 5CB and E7 far enough below $T_{NI}$, the pretilt $\theta$ has been estimated with a precision better than $1^{\circ}$.

\begin{figure}
\begin{center}
{\bf (a)}\includegraphics[width=23pc]{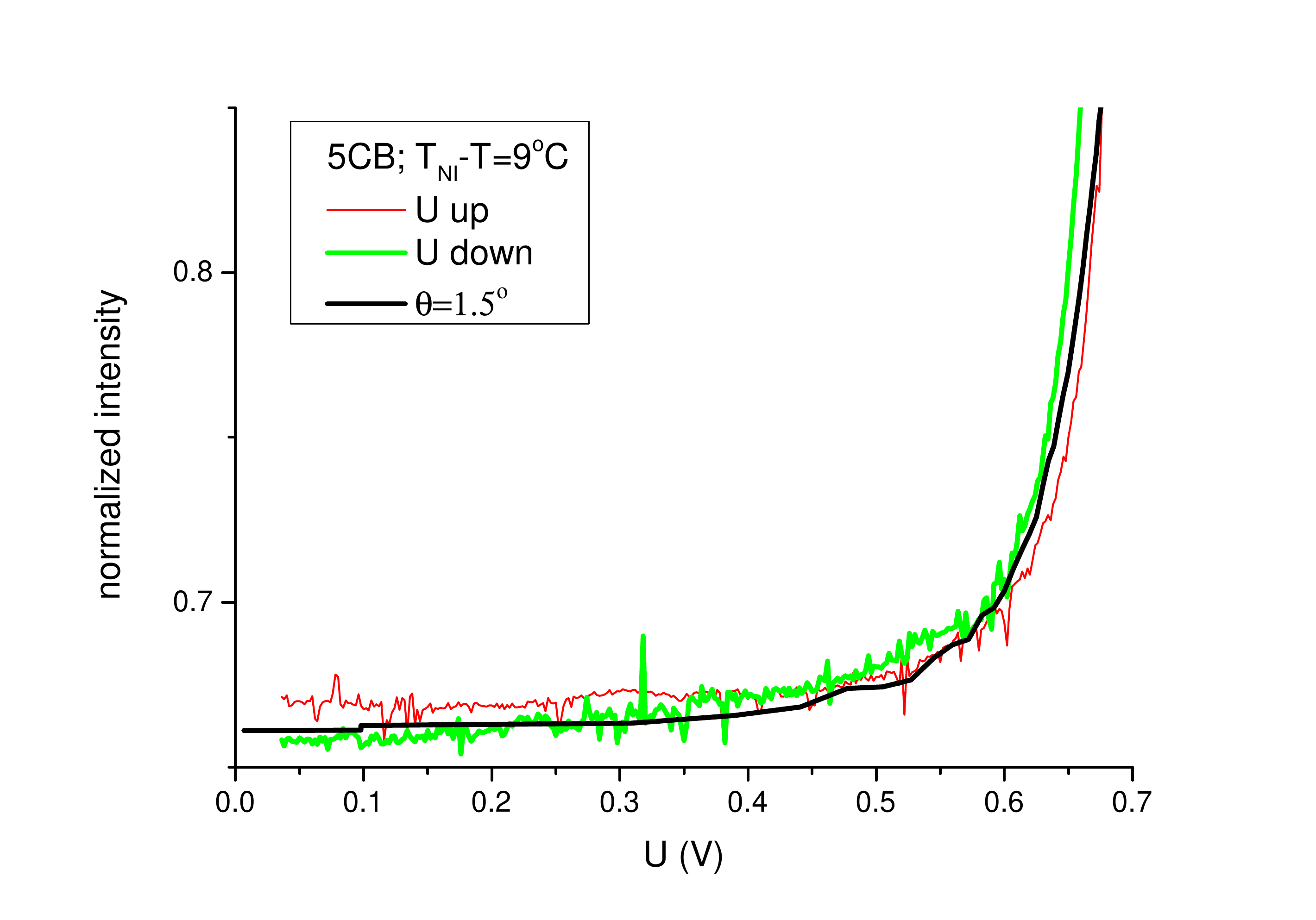} {\bf (b)}\includegraphics[width=23pc]{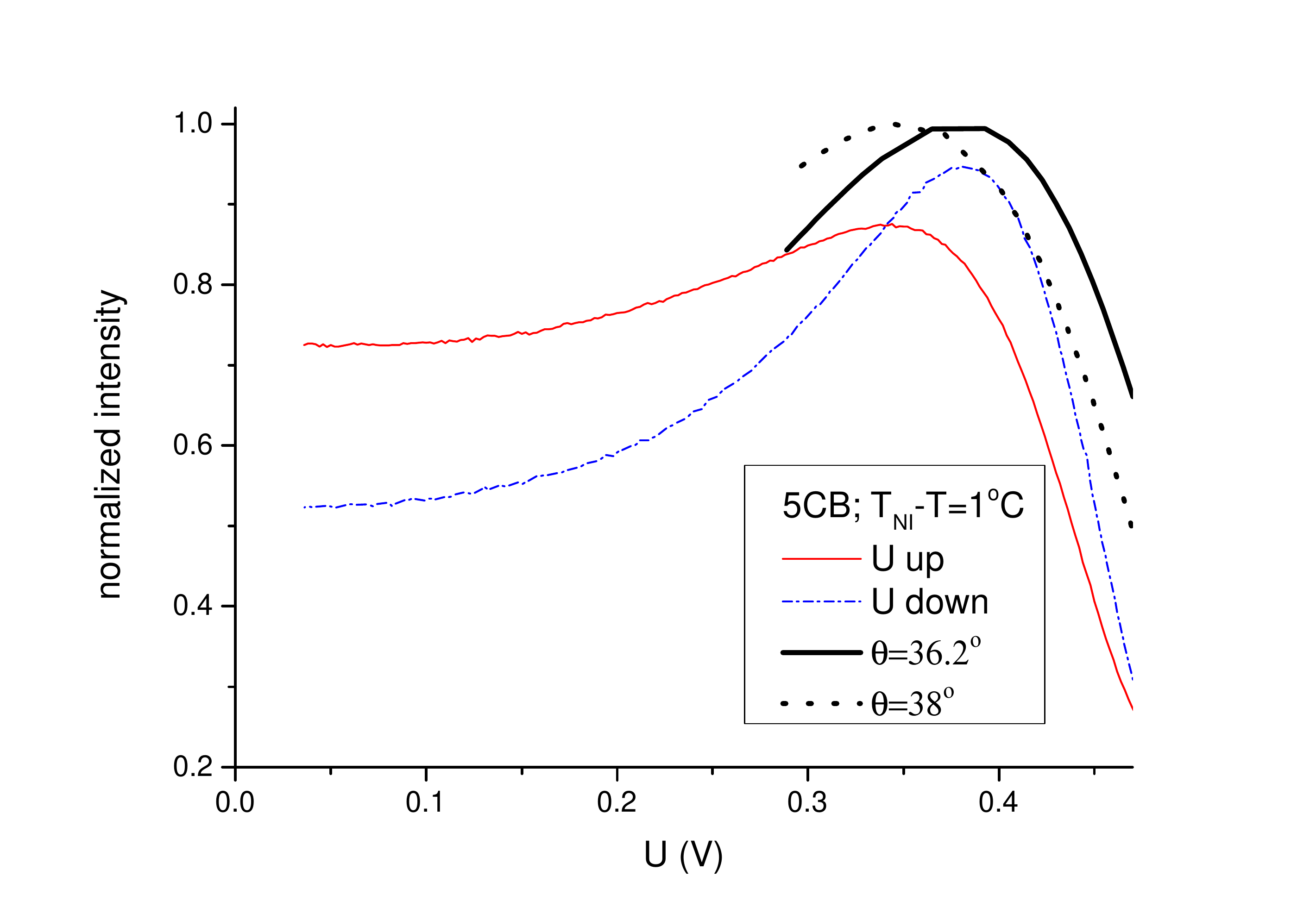}
\caption{Voltage dependence of the normalized light intensity $I$ below $U_{F}$ measured in 5CB and the calculated best fits to the experimental data at two different temperatures $\Delta T = 9^{\circ}$C ($U_{F}=0.72$V) {\bf (a)} and $\Delta T = 1^{\circ}$C ($U_{F}=0.61$V) {\bf (b)}.} \label{Toth_Fig5}
\end{center}
\end{figure}

In Fig. \ref{Toth_Fig5} the voltage dependence of the normalized light intensity $I$ is given below $U_{F}$, measured in 5CB together with the calculated best fits to the experimental data at two different temperatures (a) $\Delta T = 9^{\circ}$C and (b) $\Delta T = 1^{\circ}$C. The figure well illustrates the differences at low (far below $T_{NI}$) and high temperatures (close to $T_{NI}$).

At low temperatures [Fig. \ref{Toth_Fig5}(a)] experimental data with increasing and decreasing voltage steps ($U$ up and $U$ down, respectively) agree with each other, and they can be fitted with a single calculated curve with a well defined $\theta=1.5^{\circ}$.
At high temperatures, close to $T_{NI}$, however, several differences occur as demonstrated in Fig. \ref{Toth_Fig5}(b). First, the electric-field induced reorientation becomes completely thresholdless. Second, the experimental data with increasing and decreasing voltage steps differ substantially -- {\it cf.} $U$ up and $U$ down curves in Fig. \ref{Toth_Fig5}(b), and they can not be fitted with a single calculated curve with a given value of $\theta$. Third, in this temperature range only few oscillations have occurred as the result of the director reorientation, however, with considerably different amplitude. Consequently, the normalization of the experimental data is not straightforward [not all of the extremal values reach $0$ or $1$ after normalization -- see Fig. \ref{Toth_Fig5}(b)]. These conditions have occurred in the temperature range of $0 < \Delta T \leq 2^{\circ}$C for 5CB, and $0 < \Delta T < 5^{\circ}$C for E7. In this temperature range the pretilt $\theta$ could not be determined with such a high precision as in the lower temperature range. Taking into account the total phase shift
$\Delta \Phi$, fitting the calculated $I(U)$ curves to the first experimental extremum, and doing this separately for the increasing and decreasing voltage steps has led, however, to the determination of $\theta$ with acceptable precision -- see e.g., Fig. \ref{Toth_Fig5}(b).

\begin{figure}
\begin{center}
\includegraphics[width=23pc]{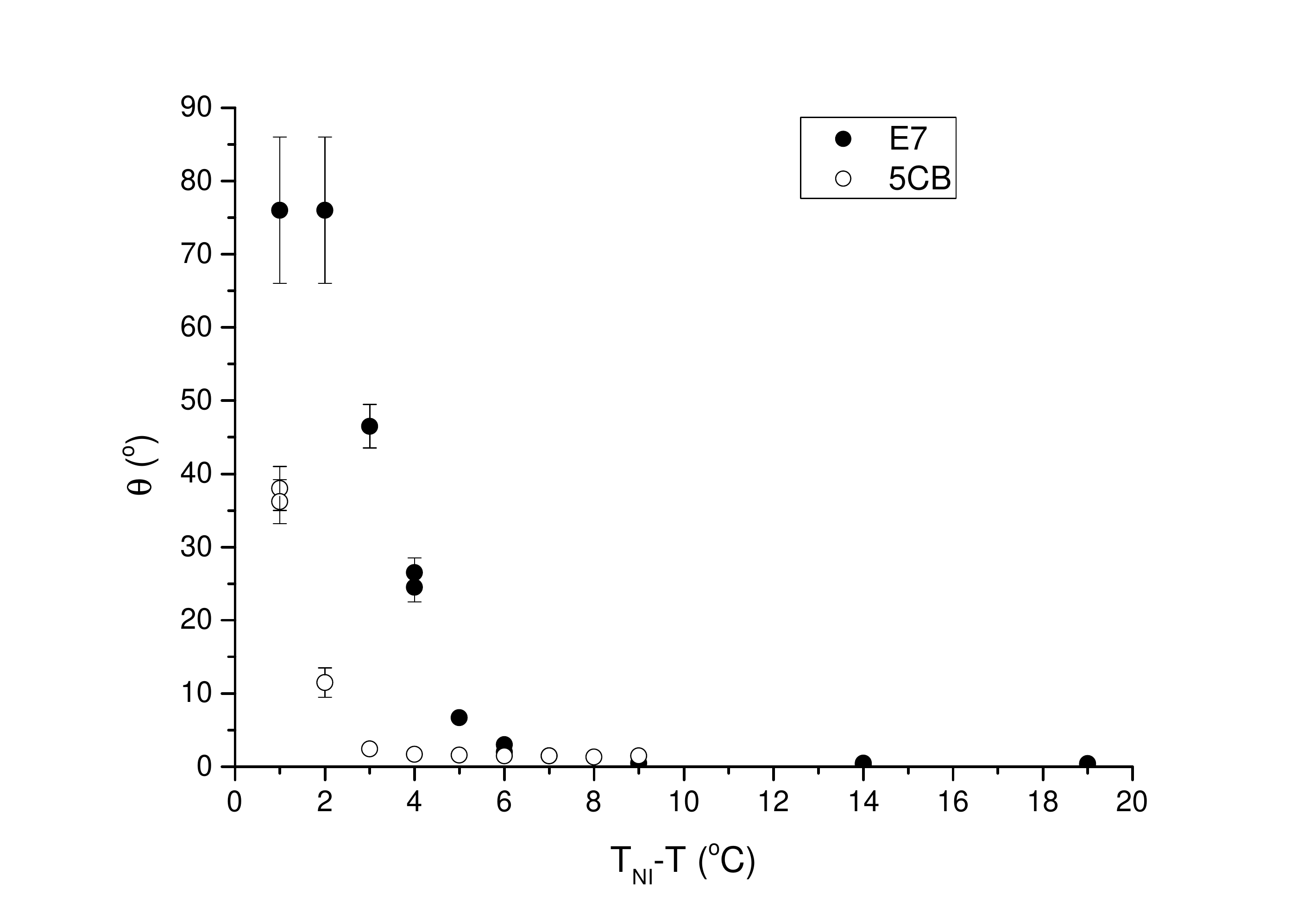}
\caption{Temperature dependence of the pretilt angle $\theta$ at the pDR1 substrate for 5CB and E7.} \label{Toth_Fig6}
\end{center}
\end{figure}

The temperature dependence of the pretilt angle $\theta$ at the pDR1 substrate, determined as described above is presented in Fig. \ref{Toth_Fig6}. As one sees, at low temperatures $\theta$ remains small (below $2^{\circ}$) providing a planar initial alignment of the LC cells. At higher temperatures ($\Delta T < 3^{\circ}$C for 5CB, and $\Delta T < 6^{\circ}$C for E7), however, $\theta$ starts to increase rapidly, and for E7 almost a complete temperature induced orientational transition from planar to homeotropic is observed just below $T_{NI}$. One has also to note that the relative temperatures at which $\theta$ starts to grow differ substantially for E7 and 5CB, indicating a subtle interaction between pDR1 and the interfacing LC.

\subsection{Zenithal (out-of-plane) photoalignment}

Next, we turn to the question how the photo-induced reorientation takes place in the high temperature range of the nematic phase, where no azimuthal photo-reorientation is observed (see Fig.~\ref{Toth_Fig2}). In accordance with the general rule, stating that after the irradiation the director becomes perpendicular to the light polarization direction, one naturally expects that an out-of-plane photoalignment takes place in that case, instead of the in-plane photoalignment. To test this expectation, the experimental setup for measuring the zenithal reorientation was used.
With this setup, if a significant out-of-plane photoalignment occurs, oscillations in the transmitted light intensity of the probe beam should appear, similarly to the measurements on the electric-, or magnetic-field induced Fr\'eedericksz transition -- see e.g., Ref. \cite{Parshin2014}.
Measurements have been performed both with the polarization of the pump beam parallel with-, and perpendicular to the initial {\bf n}. In the latter case no azimuthal photoalignment should occur. In the high temperature range of the nematic phase, where $\varphi \approx 0$, we have obtained identical temporal response of the transmitted light intensity for both polarization settings of the pump beam. In the low temperature range, where $\varphi >> 0$, the contribution of the azimuthal (twist) deformation has slightly modified the detected response for the light polarization parallel with {\bf n} compared to the case of the perpendicular light polarization.

\begin{figure}
\begin{center}
\includegraphics[width=23pc]{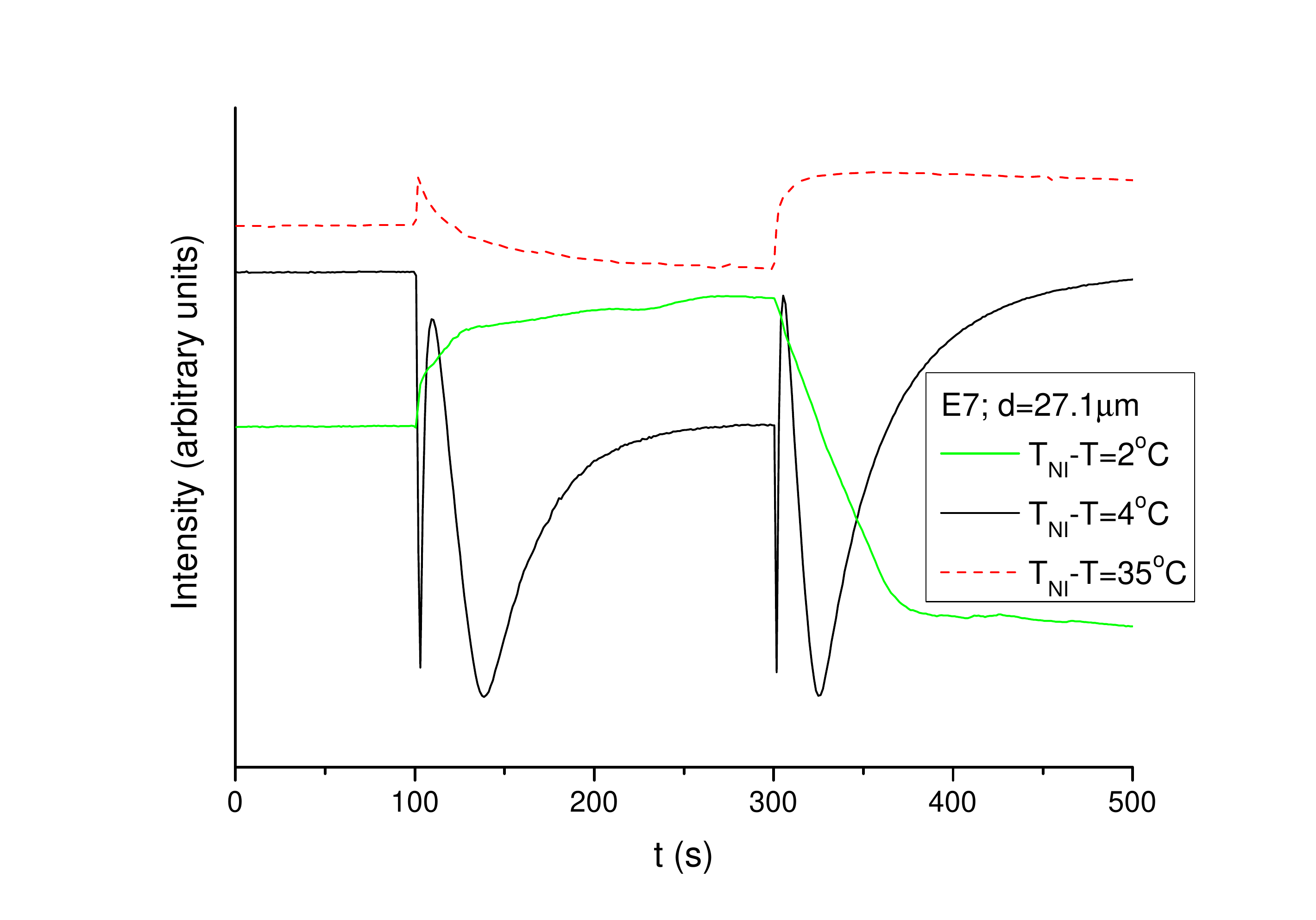}
\caption{Temporal variation of the transmitted light intensity of the probe beam measured in an E7 cell at different temperatures in the setup for detection of zenithal photo-reorientation (pump beam polarization perpendicular to {\bf n}, probe beam polarization encloses $45^{\circ}$ with {\bf n}).} \label{Toth_Fig7}
\end{center}
\end{figure}

In Fig. \ref{Toth_Fig7} we give the temporal variation of the light intensity in a $d=27.1\mu$m thick cell filled with E7 for three temperatures $\Delta T=T_{NI}-T=35^{\circ}$C, $4^{\circ}$C, and $2^{\circ}$C, for simplicity all measured with the pump beam polarization perpendicular to {\bf n}, when no azimuthal photoalignment takes place. The pump beam has been switched on at $t=100$s, and switched off at $t=300$s. At low temperature ($\Delta T=35^{\circ}$C) a slight change in the intensity has only been observed, which may originate either from a small misalignment of the director at the two bounding surfaces, or from a small misalignment of the polarization direction of the pump beam and {\bf n}, or from a slight zenithal photoalignment. At high temperature ($\Delta T=4^{\circ}$C), the switching on and off the pump beam is immediately followed by oscillations in the light intensity indicating a significant zenithal (out-of-plane) photoalignment. Interestingly, at even higher temperature, close to $T_{NI}$ ($\Delta T=2^{\circ}$C), oscillations in the light intensity disappear and only a moderate intensity change is detected indicating that the zenithal photoalignment is much smaller than that at $\Delta T=4^{\circ}$C. The reason for this can be understood by considering the temperature dependence of the pretilt angle $\theta$ at the photosensitive substrate shown in Fig. \ref{Toth_Fig6}: at $\Delta T=2^{\circ}$C the temperature induced anchoring transition has resulted in an almost homeotropic alignment $\theta \approx 75^{\circ}$, and therefore, no significant zenithal photoalignment can take place.

\begin{figure}
\begin{center}
\includegraphics[width=23pc]{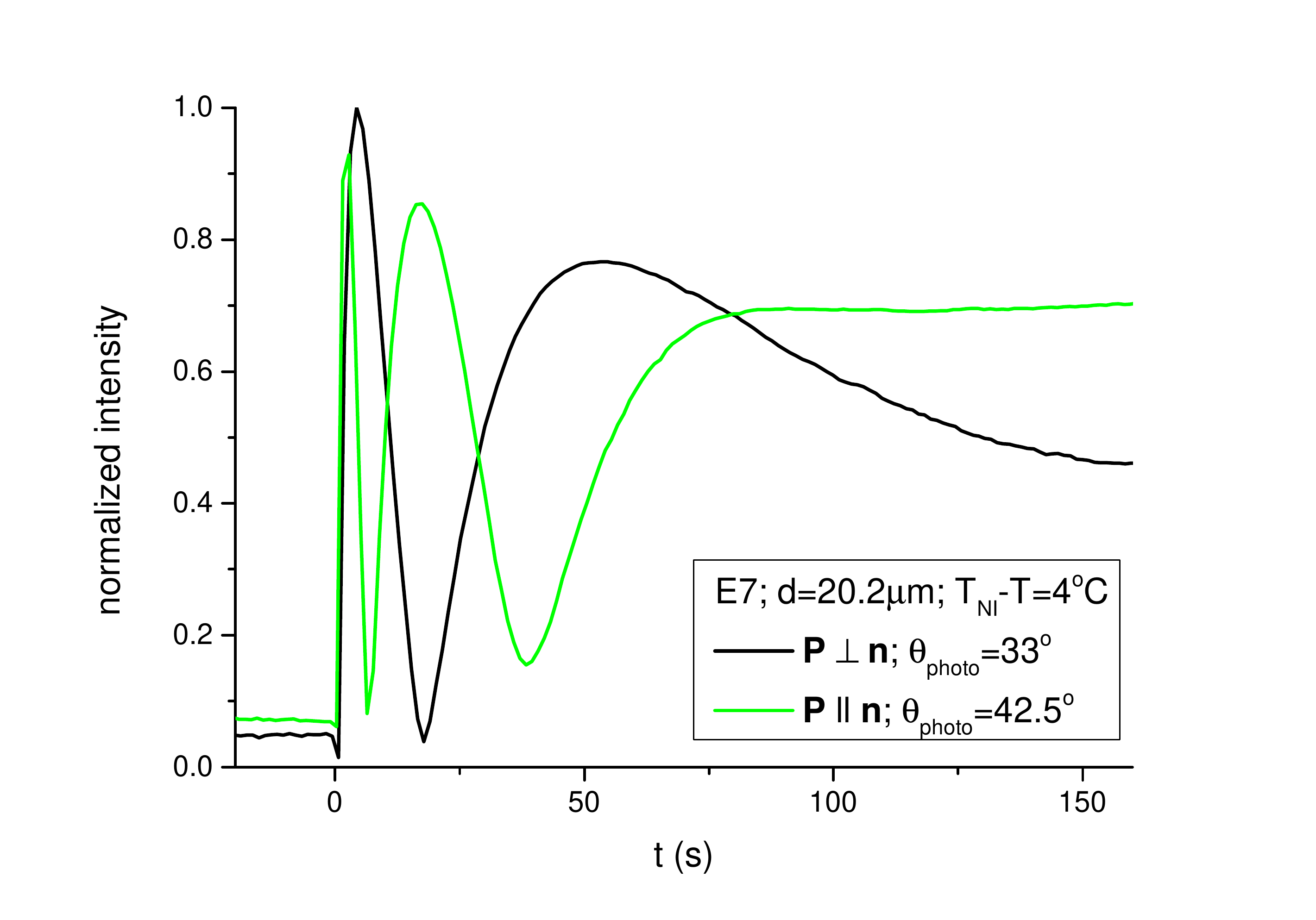}
\caption{Temporal variation of the normalized light intensity indicating a zenithal photoalignment in an E7 sample at $T_{NI}-T=4^{\circ}C$, with a pump beam of polarization ${\bf P}\perp {\bf n}$ and ${\bf P}\parallel {\bf n}$ switched on at $t=0$.} \label{Toth_Fig8}
\end{center}
\end{figure}

Though, no systematic thickness dependence of the azimuthal photo-reorientation angle $\varphi$ has been found, the temperature dependence $\varphi (T)$ varied from sample to sample considerably, as illustrated in Fig. \ref{Toth_Fig2} for E7. It is reasonable to assume a similar variation of the pretilt angle $\theta$, and of the zenithal photo-reorientation angle $\theta_{photo}$. Therefore, for a more quantitative estimation of the zenithal photoalignment angle $\theta_{photo}$ one has to use the sample for which the temperature dependence of the pretilt angle $\theta$ is determined. In Fig. \ref{Toth_Fig8} we present the temporal evolution of the probe beam normalized intensity measured in the cell filled with E7 at $T_{NI}-T=4^{\circ}$C (where a pretilt of $\theta \approx 25.5^{\circ}$ has been determined -- see Fig. \ref{Toth_Fig6}), with the pump beam of polarization ${\bf P}\perp {\bf n}$ and ${\bf P}\parallel {\bf n}$ switched on at $t=0$.
From the intensity variations zenithal photoalignment angles of $\theta_{photo} = 33^{\circ}$ and $\theta_{photo} = 42.5^{\circ}$ have been calculated for ${\bf P}\perp {\bf n}$ and ${\bf P}\parallel {\bf n}$, respectively. These values provide an estimate for the total zenithal deformation angle $\theta + \theta_{photo}$ in the range from $56^{\circ}$ to $81^{\circ}$ at this particular temperature.

\section{Discussion}
\label{Discuss}

The temperature dependence of the dynamics of azimuthal photoalignment and relaxation processes [the best illustrated in Fig.~\ref{Toth_Fig1}(b)] can be explained with the temperature dependence of the trans-cis (E/Z) isomerization. At lower temperatures, the equilibrium concentration of the trans-conformers is much higher than that of the cis-conformers, and the trans-to-cis isomerization (upon excitation) is much faster than the cis-to-trans relaxation process (when the excitation is off). With the increase of the temperature, the equilibrium ratio of the two conformers slightly change in favor of the cis isomer, the trans-to-cis isomerization somewhat slows down, while the cis-to-trans relaxation speeds up. These processes result in the observed slower photoalignment and faster relaxation at higher temperatures.

It is much more difficult to understand, however, the observed decrease of the saturation value of the azimuthal reorientation angle presented in Figs. \ref{Toth_Fig1} and \ref{Toth_Fig2}. First, even in samples with E7 and E63, in which a significant azimuthal photoalignment is detected at low temperatures, the photoinduced twist angle $\varphi$ has remained slightly below the expected value of $90^{\circ}$. In principle, this result can be interpreted by a small misalignment in the experimental setup (that of the {\bf n} at the two bounding surfaces, and/or that between the pump beam polarization and {\bf n}). However, it is very unlikely that all investigated samples were misaligned in a similar way, which always resulted in $\varphi < 90^{\circ}$. A more probable cause of the incomplete azimuthal photoalignment could be the finite zenithal anchoring strength at the pDR1 nematic LC interface as it was proposed in our recent work \cite{Janossy2018}, allowing a slight zenithal tilt that prevents the complete azimuthal photo-reorientation.

Second, the drastic decrease of the azimuthal photoalignment occurring at quite different temperatures (both on the absolute scale and on the one relative to $T_{NI}$) for different nematic LCs can not be explained by the temperature dependence of individual properties of pDR1 and LCs separately. Namely, most of the decrease in $\varphi$ occurs in the temperature range of $48-53^{\circ}$C, $35-40^{\circ}$C, and $\leq 24^{\circ}$C for E63, E7, and 5CB, respectively, while the glass transition temperature $T_g$ for pDR1 has been found by DSC measurements in the much higher temperature range of $110-130^{\circ}$C. On the other hand, LCs investigated in this work are all cyano-biphenyl based compounds/mixtures differing only in the length of their alkyl-chain, thus their physical properties are not expected to differ significantly on the temperature scale relative to $T_{NI}$. This expectation can be easily confirmed for example, by comparing the temperature dependencies of the refraction indices, the elastic constants, and the dielectric permittivities on the scale relative to $T_{NI}$ for 5CB and E7 from the literature data that we have used in our numerical calculations \cite{Raynes1979,Bogi2001,Bradshaw1985,Li2005,Karat1976}. In contrast to these similarities, the decrease of $\varphi$ occurs at quite different values of $\Delta T=T_{NI}-T$ as shown in Fig.~\ref{Toth_Fig2}.

Third, our measurements on the estimation of the pretilt angle $\theta$ at the interface between pDR1 and the nematic LCs E7 and 5CB have clearly indicated a temperature induced anchoring transition from planar towards the homeotropic alignment at the interface just below $T_{NI}$ which has further implications. Taking into account the temperature dependence of the pretilt angle shown in Fig.~\ref{Toth_Fig6}, it becomes clear why no sign of a significant zenithal photoalignment has been observed for E7 at $\Delta T = 2^{\circ}$C in Fig.~\ref{Toth_Fig7}: at that temperature the initial orientation {\bf n} is already almost homeotropic at the interface with pDR1, thus no significant zenithal photoalignment can occur. Next, if comparing the temperatures $\Delta T$, where the azimuthal photoalignment angle $\varphi$ starts to decrease (Fig.~\ref{Toth_Fig2}), and where the pretilt angle $\theta$, i.e., the initial zenithal tilt starts to increase (Fig.~\ref{Toth_Fig6}), a significant mismatch is found. Namely, the temperature induced anchoring transition starts to occur at much higher temperatures ($\Delta T < 3^{\circ}$C for 5CB, and $\Delta T \leq 6^{\circ}$C for E7) from the temperatures where the azimuthal photoalignment efficiency decreases by about $90\%$ ($\Delta T > 10^{\circ}$C for 5CB, and $\Delta T \geq 20^{\circ}$C for E7). Consequently, the temperature dependence of the pretilt angle (the temperature induced anchoring transition) can not be related directly to the disappearance of azimuthal photoalignment at high temperatures.

However, according to our interpretation, the two phenomena (the anchoring transition, and the disappearance of the azimuthal photoalignment) are closely related, and have common origin. Namely, we interpret our observations with the temperature dependence of the zenithal anchoring strength at the interface of pDR1 with the nematic LCs. We propose that the zenithal anchoring strength weakens with the increase of the temperature much faster than the azimuthal anchoring strength. Therefore, at certain temperature the zenithal anchoring strength becomes weaker than the azimuthal one, and the out-of-plane photoalignement gets energetically favorable upon irradiation. With further increase of the temperature, just below $T_{NI}$ the zenithal anchoring strength becomes so weak that thermal fluctuations trigger an anchoring transition from initially planar alignment towards the homeotropic one, even in the absence of the  irradiation. Most remarkably, the temperatures where these phenomena occur do not depend alone on the individual properties of pDR1, nor on those of the interfacing LCs exclusively. According to the presented results, the pDR1 layer senses the interfacing LC, and vice versa, i.e., the interactions between the two media have to be taken into account for the full description of the photoalignment process.

\section{Conclusions and outlook}
\label{Concl}

We found that at temperatures far from the nematic‒isotropic phase transition of the liquid
crystals, the photoalignment takes place mostly in the usual way (i.e., azimuthal reorientation of the surface director
occurs). As the temperature is raised, the behaviour of photoalignment becomes anomalous, i.e., at elevated
temperatures, the azimuthal reorientation becomes more and more incomplete; at a critical temperature, it disappears completely
and instead, the planar-to-homeotropic transition takes place. Experiments
have shown that this critical temperature for different LCs is not the same neither on the absolute scale, nor on the scale of $\Delta T = T_{NI}-T$, relative to the clearing point of the nematics. It seems to be determined rather by the interactions between the polymer and LC, i.e., by the temperature dependence of the azimuthal and zenithal anchoring strengths, the determination of which is one of the future tasks. Other, for pDR1 -- LC system relevant future works could include, e.g., {\bf (i.)} to replace the interfacing cyano-biphenyl based LC with other type of nematic(s);  {\bf (ii.)} to systematically vary the thickness of the pDR1 layer, {\bf (iii.)} to change the surface density of the azo-moieties in the pDR1 layer.

Nonetheless, the experiments presented here have evidently shown that photoalignment at the photosensitive pDR1 interfacing a nematic LC can not be considered as a two-dimensional process.
For the full description of the mechanism a more complex, three-dimensional model is needed that includes both the azimuthal and the zenithal photoalignment, takes into the account the coupling between the polymer layer and the LC (anchoring strengths), as well as the role of the pretilt angle $\theta$ (which can also be temperature dependent as in our case).

To our opinion, the observed photoalignment effects are more general, they are not restricted to the pDR1 -- LC interface only. In principle, these effects can take place at numerous other photosensitive polymer -- LC interfaces. The photoalignment process in such systems needs to be revisited, and its description needs to be extended to three-dimensions.

Moreover, the role of the coupling between the polymer and the interfacing LC (determining the azimuthal and zenithal anchoring strengths) is not restricted to photosensitive polymers.
There are numerous polymer -- LC interfaces at which the anchoring transition is triggered by other means. A representative example of such systems is the perfluoropolymer CYTOP (Asahi Glass Co. Ltd.) interfacing with the LC 4’-butyl-4-heptyl-bicyclohexyl-4-carbonitrile (CCN-47, Merck), which exhibits thermally induced homeotropic-to-planar anchoring transition around the temperature $50^{\circ}$C \cite{Uehara2014,Rasna2014}. According to recent investigations, when CYTOP interfaces other type of nematic LC, the temperature of the anchoring transition is substantially different \cite{Salamon2019}. This observation also supports our idea about the importance of the coupling between the polymer and the LC layer.

\section*{Acknowledgements}

The polymer pDR1 was kindly provided by T. K\'osa and L. Sukhomlinova (Alphamicron Inc., Kent, OH, USA). Financial support from the National Research, Development and
Innovation Office (NKFIH) Grant No. FK 125134 is gratefully
acknowledged.





\end{document}